\documentclass[twocolumn,showpacs,preprintnumbers]{revtex4}
\usepackage[dvips]{graphicx}

\begin{document}


\textbf{Polydomain ferroelectricity in very thin films [D.J. Kim \textit{et
al}., Phys.~Rev.~Lett. 95, 237602 (2005)]} 


T.W. Noh and his group (Kim \textit{et al.}\cite{KimL05}) have
recently published a seminal experimental study of very thin
ferroelectric (FE) BaTiO$_{3}$ capacitors.
They found an evidence that all their samples, including the thinnest one ($%
d=5$nm), are multi-domain (MD). Under a high enough external field,
the MD state goes over into polarization saturated state, perhaps a
single domain (SD) one, which relaxes back due to domain wall motion
if the external field $E_0$ is reduced below a certain value
$E_0=E_{0r}$. Kim \textit{et al}. have approximately identified this
field as a depolarizing field due to incomplete screening by
electrodes and claimed that it coincides with the one estimated from
electrostatics. We found that such an interpretation does not apply
and there is actually a very instructive disagreement between the
theory and experiment.

Kim \textit{et al}. have presented the polarization $P(E_{0})$ in their
Fig.~1a. The linear part of the curve has been attributed to a SD state. To
find out what the authors call a ``spontaneous polarization'', they used
linear extrapolation of $P_{0}=P(E_{0})$ to $E_{0}=0$. In fact, the
spontaneous polarization ($P_{s}$) should be defined as the polarization
when the electric field \emph{in the FE} is zero, $E_{f}=0$ ($E_{f}<0$ at $%
E_{0}=0$ because of incomplete screening by the electrodes). The value of $%
P_{s}$ has to be calculated from their data with the use of the formalism
\cite{BLcond06} and the LGD theory coefficients found in e.g. \cite
{LiCross05}. The dependence of $P_{s}$ on $d$ proves to be weak, Fig.~1a
(inset).

The idea of Kim \textit{et al}. is that the polarization relaxation sets in
(because of domains) when $E_f$ becomes opposite to the polarization. Let us
find the corresponding external field $E_0=E_{0b}$. Since $%
E_{f}=E_{0}-2\lambda P/\left( \epsilon _{0}\epsilon _{e}d\right) $
\cite {BLcond06}, then $E_{0b}=2\lambda P_{s}/\left( \epsilon
_{0}\epsilon _{e}d\right) $ with $\lambda=0.8${\AA}
($\epsilon_e$=8.45) the screening length (dielectric constant) of
the electrodes\cite{KimL05}. This field is $\sim 60$\% larger than
that determined by Kim
\textit{et al.} as the external field $E_0=E_{0r}$ for $d=5$nm (raw data), while at $%
d=30$nm the difference practically vanishes, Fig.~1a. This means
that while in relatively thick films the domains begin to appear at
small fields opposing polarization\cite{Merz}, in the thinnest film
($5$nm) the field is quite large, estimated as $-\left( 490\pm
70\right)$kV/cm \cite{BLcond06}, and yet there is no polarization
relaxation for at least $10^{3}$s \cite {KimL05}.
\begin{figure}[tbp]
\includegraphics{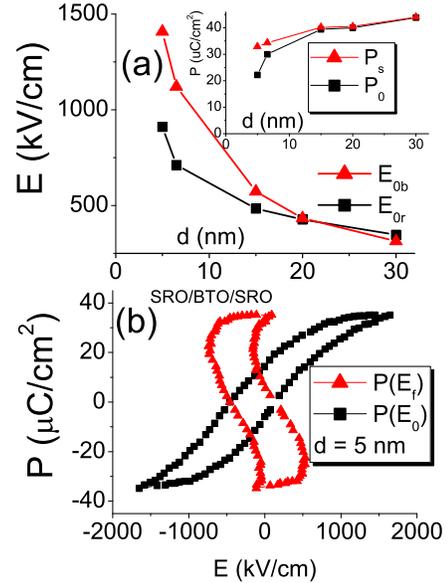}
\vskip -0.5mm
\caption{ (a) The external field $E_{0b}$ (where $E_f=0$) and $E_{0r}$ where
the relaxation of polarization starts\protect\cite{KimL05}. Inset: the
spontaneous polarization $P_s$ and the extrapolated $P_0$\protect\cite
{KimL05}. (b) The measured $P(E_0)$ and the ``actual" $P(E_f)$ hysteresis
loops. }
\label{fig:fig1}
\end{figure}
At much shorter time intervals, $\sim 10^{-3}$s, the field is
apparently present too (see below). This is qualitatively similar to
a dependence of the Merz's activation field  on an application
(switching) time and a film thickness, although the thickness
dependence of the field at short application times is not captured
by Merz's empirical formula\cite{Merz}.

The hysteresis loop $P=P(E_{f})$ has an unusual \emph{negative
slope}, Fig.~1 (raw data for $2$kHz, $d=5$nm, Ref.~2 in
\cite{KimL05}).  The negative slope is a hallmark of MD state and
was predicted for an ideal FE film in a capacitor with a voltage
drop across a ``dead" layer near FE/electrode interface\cite{BL01}.
There are no dead layers in the present system but the same effect
is at work because of a voltage drop across a screening layer in the
electrode with the thickness $\lambda$. Even numerically, the
observed negative slope is surprisingly close to the theoretical
value \cite{BLcond06}. This means that in the thinnest films the
behavior of the domain structure may be determined by the
electrostatics rather than by pinning etc. If the value of the field
promoting the domain formation is defined by the thickness only and
not by properties of electrodes or an electrode-film interface, one
can speculate about the properties of electrodes which can
facilitate a smaller field in short-circuited sample and a longer
retention of a SD state. We have found
that for $d=5$nm such an electrode should have $\lambda /\epsilon _{e}<0.03$%
\AA. Since in \cite{KimL05} this value is about $0.1$\AA, it does
not seem impossible to find such an electrode. Alternatively,
thinner films may show longer retention. We thank T.W. Noh and his
group for kindly sharing their data and many useful discussions. APL
is partially supported by MAT2003-02600 and S-0505/MAT/000194.

\textbf{A.M. Bratkovsky$^{1}$ and A.P. Levanyuk$^{1,2}$}

$^{1}$Hewlett-Packard Laboratories, Palo Alto, California

$^{2}$Universidad Aut\'{o}noma de Madrid, 28049 Madrid, Spain


\end{document}